\documentclass[fleqn,usenatbib]{mnras}

\usepackage{newtxtext,newtxmath}

\usepackage[T1]{fontenc}
\usepackage{ae,aecompl}
\usepackage{graphicx}	
\usepackage{amsmath}	
\usepackage{amssymb}	
\usepackage{threeparttable}
\usepackage{threeparttablex}
\usepackage{longtable}
\usepackage{tabularx}
\usepackage{subfigure}
\usepackage{rotating}
\usepackage{enumerate}
\usepackage{framed}
\usepackage[T1]{fontenc}
\usepackage{aecompl}
\usepackage{soul}
\usepackage{ltablex}
\usepackage{pdflscape}
\usepackage{xtab}
\usepackage{afterpage}
\usepackage{setspace}
\usepackage[utf8]{inputenc}
\usepackage[english]{babel}
\usepackage[dvipsnames]{xcolor}
\usepackage[normalem]{ulem}

\newcommand{\note}[1]{\textbf{\textcolor{red}{{\fontfamily{garamond}\selectfont{#1}}}}}
\newcommand{\lclash}{Cluster Lensing And Supernova survey with Hubble}
\newcommand{\sbu}{mag arcsec$^{-2}$}

\newcommand{\hst}{\textit{HST}}
\newcommand{\chandra}{\textit{Chandra}}
\newcommand{\XMM}{\textit{XMM-Newton}}

\newcommand{\mfive}{M$_{500,c}$}
\newcommand{\mtwo}{M$_{200,c}$}
\newcommand{\rfive}{r$_{500,c}$}

\newcommand{\red}{F160W}

\newcommand{\til}{$\thicksim$}

\newcommand{\Msun}{M$_\odot$}
\newcommand{\Mstell}{M$_{\star}$}

\newcommand{\tenthirteen}{$\times 10^{13}$}
\newcommand{\tenfourteen}{$\times 10^{14}$}

\newcommand{\logmfive}{log(M$_{500,c}$ [M$_\odot$])}
\newcommand{\logmstell}{log(M$_{\star}$/M$_\odot$)}
\newcommand{\logmfivenorm}{log(M$_{500,c}$/2\tenfourteen \Msun)}
\newcommand{\logMstell}{log(M$_{\star}$/M$_\odot$)}
\newcommand{\bcgicl}{BCG+ICL}
\newcommand{\programs}{\#13677 \& \#14327}
\newcommand{\sz}{Sunyaev-Zel'dovich}
\newcommand{\xdcp}{XDCP J0044.0$-$2033}
\newcommand{\sparcszerotwo}{SpARCS-J0224}
\newcommand{\sparcszerothree}{SpARCS-J0330}
\newcommand{\sparcsonezero}{SpARCS-J1049}
\newcommand{\idcs}{IDCS J1426.5$+$3508}
\defcitealias{Gonzalez2005}{GZZ05}
\defcitealias{Gonzalez2013a}{GZZ13}
\defcitealias{DeMaio2015}{Paper I}
\defcitealias{DeMaio2018}{Paper II}
\defcitealias{BC03}{BC03}

\title[BCG+ICL Growth Over the Past 10 Gyrs]
{The Growth of Brightest Cluster Galaxies and Intracluster Light Over the Past Ten Billion Years}

\author[T. DeMaio et al.]{Tahlia DeMaio,$^{1}$
Anthony H. Gonzalez,$^{1}$
Ann Zabludoff,$^{2}$
Dennis Zaritsky,$^{2}$
\newauthor
Greg Aldering,$^{3}$
Mark Brodwin,$^{4}$
Thomas Connor,$^{5}$
Megan Donahue,$^{6}$
\newauthor 
Brian Hayden,$^{7}$
John S. Mulchaey,$^{5}$
Saul Perlmutter,$^{3,8}$
and S. A. Stanford$^{9}$ 
\\
$^{1}$Department of Astronomy, University of Florida, Gainesville, FL 32611\\
$^{2}$Department of Astronomy, University of Arizona, Steward Observatory, Tucson, AZ  85721\\
$^{3}$Lawrence Berkeley National Laboratory, 1 Cyclotron Road, MS 50B-4206, Berkeley, CA 94720, USA\\
$^{4}$Department of Physics and Astronomy, University of Missouri, 5110 Rockhill Road, Kansas City, MO 64110, USA\\
$^{5}$The Observatories of the Carnegie Institution for Science, 813 Santa Barbara St, Pasadena, CA 91101\\
$^{6}$Department of Physics and Astronomy, Michigan State University, East Lansing, MI 48824\\
$^{7}$Space Telescope Science Institute, 3700 San Martin Drive, Baltimore, MD 21218, USA\\
$^{8}$Astronomy Department, University of California at Berkeley, Berkeley, CA 94720, USA\\
$^{9}$Department of Physics, University of California, One Shields Avenue, Davis, CA 95616, USA
}
\date{Accepted form publication in \mnras.}

\pubyear{2018}

\begin{document}
\label{firstpage}
\pagerange{\pageref{firstpage}--\pageref{lastpage}}
\maketitle

\begin{abstract}
We constrain the evolution of the brightest cluster galaxy plus intracluster light (\bcgicl) using an ensemble of 42 galaxy groups and clusters that span redshifts of $z=0.05-1.75$ and masses of \mfive$=2\times10^{13}-10^{15}$ \Msun.  Specifically, we measure the relationship between the \bcgicl\ stellar mass \Mstell\ and \mfive\ at projected radii $10<r<100$~kpc for three different epochs. At intermediate redshift ($\bar{z}=0.40$), where we have the best data, we find \Mstell$\propto$\mfive$^{0.48\pm0.06}$. Fixing the exponent of this power law for all redshifts, we constrain the normalization of this relation to be $2.08\pm0.21$ times higher at 
$\bar{z}=0.40$ than at high redshift ($\bar{z}=1.55$).
We find no change in the relation from intermediate to low redshift ($\bar{z}=0.10$). In other words, for fixed \mfive, \Mstell\ at $10<r<100$~kpc increases from $\bar{z}=1.55$ to $\bar{z}=0.40$ and not significantly thereafter. Theoretical models predict that
the physical mass growth of the cluster from $z=1.5$ to $z=0$ within \rfive\ is 1.4$\times$, excluding evolution due to definition of \rfive. We find that \Mstell\ within the central 100~kpc increases by $\sim3.8\times$ over the same period. Thus, the growth of \Mstell\ in this central region is more than a factor of two greater than the physical mass growth of the cluster as a whole. Furthermore, the concentration of the \bcgicl\ stellar mass, defined by the ratio of stellar mass within 10 kpc to the total stellar mass within 100~kpc, decreases with increasing \mfive\ at all $z$. We interpret this result as evidence for inside-out growth of the \bcgicl\ over the past ten Gyrs, with stellar mass assembly occuring at larger radii at later times.
\end{abstract}




\section{Introduction}
In the paradigm of hierarchical assembly, massive halos are assembled from smaller halos. The most massive galaxies at any epoch represent the culmination of this process and are expected, as a class, to continue to grow from the accretion of smaller galaxies up to the current time. Brightest cluster or group galaxies  (hereafter BCGs) trace the most overdense peaks in the matter distribution. 
Connected to the evolution of the BCG is the build-up of its extended stellar halo, the so-called intracluster light (hereafter ICL), which extends to hundreds of kpc
\citep{Oemler1976, schombert1988, Gonzalez2005,Zibetti2005}. The rate of growth of the sum of these two components (hereafter \bcgicl), and the time of ICL formation relative to that of the BCG, thus represent important benchmarks for models of cluster and massive galaxy formation.

While the physics underlying predictions of massive galaxy growth 
is straightforward, comparisons of the rate of \bcgicl\ growth between models and observations, and even among the observations themselves, yield disparate results. For example, \citet{DLB2007} used semi-analytic models based upon the Millennium simulations to track the stellar mass growth of BCGs via merger trees, finding a factor of three growth since $z=1$. More recent semi-analytic models that also include the ICL \citep{Contini2013a, contini2018} find that ICL growth is delayed compared to growth of the BCG, resulting in an even larger growth of the \bcgicl\ since $z=1$. In contrast, observational studies of BCG growth typically have found less stellar mass evolution, from essentially none \citep{Whiley2008, Collins2009, Stott2010} to 1.3 to 2$\times$ over a comparable redshift range in more recent papers \citep{Lidman2012a,  LinYT2013, Bellstedt2016, Zhang2016}. A challenge in comparing the observational results is the use of different apertures
to define the BCG stellar mass \citep{Burke2000, vanDokkum2010, Zhang2016}, leading to the inclusion of more or less of the extended ICL component.  Any systematic discrepancy with theory may arise from a failure to account for ICL stars at larger radii \citep{Zhang2016}, particularly given the expected delayed growth of the ICL.

If the late-time build-up of the BCG+ICL component is such that the ICL is assembling faster than the BCG
\citep[e.g.][]{Contini2013a,contini2018}, then one will observe ``inside-out" growth in the stellar mass distribution. With observations of the radial profile of the \bcgicl\ over a range of redshifts, it is possible to determine how the concentration of the stellar mass changes over time and therefore to test the inside-out scenario.

Measurements of the \bcgicl\ radial profile are technically challenging.
and consistent from $\bar{z}=0.4$ to $\bar{z}=0.10$. While the extended halo of the \bcgicl\ often dominates the total stellar light within the central few hundred kpc \citep[e.g.][]{Gonzalez2005,Gonzalez2013a,zhang2019}, it can be of extremely low surface brightness, far below that of the background sky.
Meticulous control of systematics is thus required, even at low redshift. Cosmological dimming and a decrease in the stellar mass of the \bcgicl\ with increasing redshift add to the challenge when measuring the \bcgicl\ at higher redshift.

As a result, there are only a few measurements of the \bcgicl\  extending beyond the central $\sim20$~kpc in clusters at $z\ga1$ \citep{Burke2012a,Zhang2016,ko2018}.  \cite{Burke2012a} measured the  ICL in a sample of six clusters at $z\sim1$, finding that the stellar content with $\mu_J>22$ \sbu\ represents only $1-4$\% of the total cluster luminosity within \rfive\ and that it grows $2-4\times$ by the present day. \citet{ko2018} provides the only \hst\ study of the ICL at $z>1$, arguing for modest growth based on their detection of a substantial ICL  in MOO J1014+0038 ($z=1.24)$, which we also include in this study.

In this paper, we measure the \bcgicl\ out to projected radii of 100~kpc, where the ICL dominates \citep{Gonzalez2005,DeMaio2015}, for 42 groups and clusters spanning a range of redshifts. This work includes new \hst\ measurements of the \bcgicl\ for seven clusters at $z\ge1.24$ (mean redshift $\bar{z}=1.55$). We make no attempt to distinguish between the BCG and ICL components, constraining their sum instead. Thus, our results can be used to directly test models in which the BCG+ICL is determined within this aperture. We further quantify the \bcgicl\ stellar mass within fixed physical apertures of 10 and 100~kpc, and within an annulus of 10 to 100~kpc, and use these data to constrain the radial dependence of the growth.

These data, in combination with our previous work, provide constraints on when and how the mass of the \bcgicl\ was established as a function of cluster mass. In Section \ref{sec:piii_sample}, we describe the high redshift {\it HST} cluster sample ($\bar{z}=1.55$), which we combine with lower-redshift systems  at $\bar{z}=0.10$ and $\bar{z}=0.40$ to create a wide redshift baseline. Because these latter samples are both from our own previous work, the comparison across mass and redshift is as consistent and straightforward as possible. In Section \ref{sec:piii_reduction}, we summarize the reduction methods for the new high-redshift clusters and describe how we produce their surface brightness profiles out to \til100~kpc. We present the results of our analysis in Section \ref{sec:results}, including how the observed trend between the \bcgicl\ stellar mass and the total cluster mass evolves and how the concentration of the \bcgicl\ stellar mass evolves. We summarize our conclusions in \S\ref{sec:piii_conclusion}. We use the WMAP9 cosmology 
\citep[$H_0$=69.3 km Mpc$^{-1}$ s$^{-1}$, $\Omega_m=0.286$;][]{WMAP9} as in \citet[ hereafter Paper I]{DeMaio2015} and \citet[hereafter Paper II]{DeMaio2018}. Throughout this paper, $r$ refers to projected radius from the center of the BCG, \rfive\ is radius within which the cluster overdensity equal to 500 times the critical density of the Universe at the cluster redshift, and \mfive\ is the mass enclosed within this radius.

\section{Sample}
\label{sec:piii_sample}

At low redshifts, we use a sample of 12 clusters drawn from  \citet[hereafter GZZ05]{Gonzalez2005} for which \citet[hereafter GZZ13]{Gonzalez2013a} derive \mfive\ from \XMM\ observations. These 12 clusters span the mass range 0.9\tenfourteen\  \Msun\ $<$ \mfive $<$ 6\tenfourteen\ \Msun\ at $z<0.15$ ($\bar{z}=0.10$). \citetalias{Gonzalez2005} fit  their observed  \bcgicl\ surface brightness profiles out to $r>300$~kpc with two de Vaucouleur profiles. 
We refer the reader to their Table 4 for the best-fit parameters of each system. For these systems, we first transform the photometry from the Cousins magnitudes in \citetalias{Gonzalez2005} to Sloan $i^\prime$ using the  color correction of   \citet{Jordi2006}  for $R-I=0.5$ before computing the luminosities.\footnote{The systematic uncertainty in the mean $R-I$ color of the Landolt calibrators used in \citetalias{Gonzalez2005} is subdominant to statistical uncertainties in this analysis.}  We then use the $i^\prime$ absolute magnitudes and structural parameters to compute the \bcgicl\ luminosities within fixed physical apertures (e.g. 10~kpc, 100~kpc), enabling comparison with aperture luminosities measured for the other subsamples described below. The \mfive\ values from \citetalias{Gonzalez2013a} are derived from \XMM\ X-ray temperatures, $T_x$, using the \cite{Vikhlinin2009} prescription.

At intermediate redshift, $0.29\leq z\leq0.89$  ($\bar{z}=0.40$), we have 23 systems from \citetalias{DeMaio2018}. These clusters span a wide range in \mfive, 3\tenthirteen$-9$\tenfourteen\ \Msun. Data for clusters with \mfive above 10$^{14}$ \Msun\ are from the \lclash\  \citep[CLASH][]{p12a}, while data for the others originate from HST Program \#12575 (PI: Gonzalez). \citetalias{DeMaio2018}  provides more details on these intermediate-redshift systems. For this paper, we focus on the \red\ surface brightness profiles out to 100~kpc. We use these profiles to find the total luminosity and the stellar mass content of the \bcgicl\ in each system. 

\mfive\ values for all these systems are derived from X-ray temperatures, using the \citet{Vikhlinin2009} transformation, \mfive$=M_0(T_x/5 {\rm \ keV})^\alpha_x E(z)^{-1}$ with $M_0=3.02\pm$0.11\tenfourteen\ $h^{-1}$ \Msun\ and $\alpha_x=1.53\pm0.08$. The CLASH sample all have published \chandra\ X-ray temperatures \citep{donahue2014,donahue2016}. For the lower mass systems 
X-ray temperatures come from a mixture of \chandra\ and \XMM\ data.
There is a systematic offset for the derived cluster masses of systems with $k_B T_x>5$ keV due to calibration effects; \chandra\  masses have been found to be \til15\% higher than those determined with \XMM\ observations \citep{Mahdavi2013, Schellenberger2015}.
\footnote{
\citet{donahue2014} also provides a comparison of \chandra\ and \XMM\ masses for many of the CLASH clusters.}
To bring the \mfive\ values of both the low-redshift sample of \citetalias{Gonzalez2005} and our intermediate redshift sample to a common framework, we apply a correction factor of 15\% to the \citetalias{Gonzalez2013a} masses.  Final \mfive\ values after all corrections are presented in Table \ref{table:piii_sample}.

\begin{figure}
    \includegraphics[width=\columnwidth]{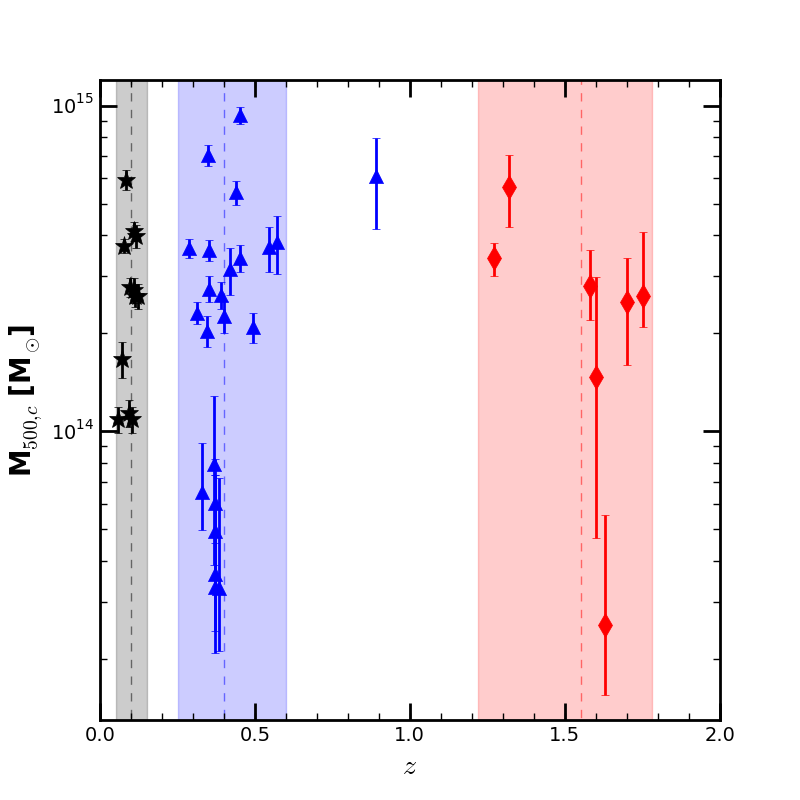}
    \caption[Cluster mass and redshift distribution of complete sample.]{Cluster mass and redshift distribution of complete sample with low-redshift sources from \citetalias{Gonzalez2005} (black stars), intermediate redshift systems from \citetalias{DeMaio2018} (blue triangles) and high-redshift clusters (red diamonds). The dashed vertical lines denote the mean redshift for each subsample, while the shaded regions indicate the redshift range of each subsample used to compute the mean redshifts. For the intermediate redshift sample, we exclude the cluster CL J1226+3332 at $z=0.89$ from subsequent analysis because its redshift is far from the mean of the sample and we lack other clusters at comparable redshift. from subsequent analysis Quoted \mfive\ values are primarily derived from the X-ray temperature mass proxy using the \cite{Vikhlinin2009} prescription. All \mfive\ values, literature sources, and mass proxies are listed in Table \ref{table:piii_sample}. 
      }
    \label{fig:piii_zm500}
\end{figure}

We extend the redshift baseline of this sample with clusters from HST Programs \programs\ (PI: Perlmutter; hereafter the high-redshift sample). 
These programs were designed to observe high-redshift supernova in massive clusters at $z>1$ to constrain the time variation of dark energy.  The clusters in the sample cover a similar mass range to the lower redshift samples (Figure \ref{fig:piii_zm500}). We use the \red\ observations of 7 clusters from this program (Table \ref{table:piii_sample}),  which all lie at $z\ge1.24$ ($\bar{z}=1.55$).
For \idcs\ we supplement the Perlmutter \hst\ data with \red\ imaging from HST Program \#12994 (PI: Gonzalez), as described in \cite{Mo2016}.

Redshifts, BCG positions and cluster masses are compiled from the literature for the high-redshift systems as follows:
\begin{enumerate}[1.]
    \item For \sparcszerothree\ and \sparcszerotwo\, we use the redshifts and cluster centers of \cite{Nantais2016}.
    For these two systems we use \mtwo\ values from \cite{Lidman2012a}.
    We convert the \mtwo\ masses to \mfive\ assuming an NFW \citep{NFW} profile with a \cite{Duffy2008a} concentration, using the Python package \texttt{colossus} \citep{colossus}.

 \vspace{1ex}
   
    \item For \sparcsonezero\, we use the BCG position as defined in \cite{Webb2015}.
    \cite{Webb2015} provide
    two estimates of this cluster's mass. With richness as a mass proxy, they find the cluster mass within 500~kpc to be (3.8$\pm1.2$)\tenfourteen\ \Msun. Using the velocity dispersion from 27 members within 1.8 Mpc they find a virial mass of (8$\pm$3)\tenthirteen\ \Msun. A more recent weak lensing analysis (Finner et al., submitted) yields \mfive$=2.5\pm0.9$\tenfourteen\ \Msun. All subsequent analysis uses this weak lensing-derived \mfive.

    \item For SPT-CL J0205$-$5829, the BCG identification and \mfive\ are based on values in Tables 1 and B1 of \cite{Chiu2016}.    

    \item For \xdcp, we identify the galaxy ID1 in \cite{Fassbender2014} as the BCG and use the \mfive\ estimate derived from the \chandra\ X-ray temperature in \cite{Tozzi2015}.  
    
    \item For MOO J1014+0038, we adopt cluster coordinates and \mfive\ 
    from \cite{Brodwin2015}. The \mfive\ measurement is based on the observed \sz\ (SZ) signal obtained with \emph{CARMA}.    
    
    \item For \idcs, we associate the BCG identified in \cite{Brodwin2016} as the cluster center. Additionally, we use the value of \mfive\ they derived using the $Y_x-$\mfive\ mass proxy. 
    
\end{enumerate}
See Table \ref{table:piii_sample} and Figure \ref{fig:piii_zm500} for cluster details of the entire composite sample, which includes redshift, \mfive, and references for the photometry and masses. 

\begin{table}
	\caption{Composite Cluster Sample}
	\begin{threeparttable}
		\begin{tabular}{lllcc}
    		\hline
            Cluster & z & M$_{500,c}$ & Notes\\
 &  & 10$^{14}$ [M$_\odot$]  & \\
\hline
\multicolumn{4}{c}{Low-redshift ($\bar{z}=0.10$)} \\
\hline
Abell 2401  & 0.0578 & 0.95$\pm0.1$ & $a$\\
Abell S0296 & 0.0699 & 1.45$\pm0.21$ & $a$\\
Abell 3112  & 0.0759 & 3.23$\pm0.19$ & $a$\\
Abell 1651  & 0.0853 & 5.15$\pm0.42$ & $a$\\
Abell 2955 & 0.0945 & 0.99$\pm0.11$ & $a$\\
Abell 4010 & 0.0963 & 2.41$\pm0.18$ & $a$\\
Abell 2984 & 0.1044 & 0.95$\pm0.1$ & $a$\\
Abell 2811 & 0.1082 & 3.59$\pm0.28$ & $a$\\
Abell S0084& 0.1087 & 2.37$\pm0.24$ & $a$\\
Abell 0122 & 0.1127 & 2.26$\pm0.19$ & $a$\\
Abell 2721 & 0.1149 & 3.46$\pm0.32$ & $a$\\
Abell 3693 & 0.1237 & 2.26$\pm0.23$ & $a$\\
\hline
\multicolumn{4}{c}{Intermediate-redshift ($\bar{z}=0.40$)$^\dagger$} \\
\hline
Abell 611       & 0.288 & 3.66$\pm0.25$ & $b$\\
MS2137$-$2353    & 0.313 & 2.31$\pm0.18$ & $b$\\
XMMXCS J022045.1$-$032555.0  & 0.33 & 0.65$_{-0.15}^{+0.27}$ & $b$\\
RX J1532$+$3021  & 0.345 & 2.04$\pm0.23$ & $b$\\
RX J2248$-$4431  & 0.348 & 7.06$\pm0.52$ & $b$\\
MACS1931$-$2635  & 0.352 & 2.75$\pm0.25$ & $b$\\
MACS1115$+$0129  & 0.352 & 3.60$\pm0.28$ & $b$\\
SG 1120$-$1202$-$4 & 0.3688 & 0.80$_{-0.41}^{+0.49}$ & $b$\\
XMMXCS J011140.3$-$453908.0  & 0.37 & 0.60$_{-0.15}^{+0.22}$ & $b$\\
SG 1120$-$1202$-$2 & 0.3704 & 0.33$_{-0.09}^{+0.15}$ & $b$\\
SG 1120$-$1202$-$1 & 0.3707 & 0.49$_{-0.14}^{+0.24}$ & $b$\\
SG 1120$-$1202$-$3 & 0.3713 & 0.36$_{-0.15}^{+0.37}$ & $b$\\
RX J1334.0$+$3750& 0.384 & 0.33$_{-0.12}^{+0.39}$ & $b$\\
MACS1720$+$3536  & 0.391 & 2.63$\pm0.24$ & $b$\\
MACS0429$-$0253  & 0.399 & 2.26$\pm0.25$ & $b$\\
MACS0416$-$2403  & 0.42 & 3.14$\pm0.51$ & $b$\\
MACS1206$-$0848  & 0.44 & 5.43$\pm0.46$ & $b$\\
MACS0329$-$0211  & 0.45 & 3.41$\pm0.33$ & $b$\\
RX J1347$-$1145  & 0.451 & 9.38$\pm0.56$ & $b$\\
MACS1311$-$0310  & 0.494 & 2.09$\pm0.22$ & $b$\\
MACS1149$+$2223  & 0.544 & 3.67$\pm0.58$ & $b$\\
MACS2129$-$0741  & 0.57 & 3.81$\pm0.78$ & $b$\\
CL J1226$+$3332 & 0.89 & 6.08$\pm1.89$ & $b$\\
\hline
\multicolumn{4}{c}{High-redshift ($\bar{z}=1.55$)} \\
\hline
MOO J1014$+$0038    & 1.24 & 3.40$\pm0.40$ &$c,d$\\
SPT-CL J0205$-$5829 & 1.32 & 5.65$\pm1.40$ &$c,e$\\
XDCP J0044.0$-$2033 & 1.58 & 2.8$_{-0.6}^{+0.8}$ &$c,f$\\
SpARCS-J0330      & 1.6  & 1.47$_{-1.0}^{+1.5}$  &$c,g$\\
SpARCS-J0224      & 1.63 & 0.26$_{-0.1}^{+0.3}$  &$c,g$\\
SpARCS-J1049      & 1.7  & 2.5$\pm0.90$ &  $c,h$\\
IDCS J1426.5$+$3508 & 1.75 & 2.6$_{-0.5}^{+1.5}$ &$c,i$\\
            \hline
		\end{tabular}
	\begin{tablenotes}
        \item $a$ Photometry from \cite{Gonzalez2005} and masses from \cite{Gonzalez2013a}, $b$ Photometry from \cite{DeMaio2018}, masses from \cite{p12a}, $c$ Imaging from HST Programs \programs, $d$ \mfive\  from \cite{Brodwin2015}, $e$ \mfive\ from \cite{Chiu2016}, $f$ \mfive\ from \cite{Tozzi2015},  $^g$ Mass converted from \mtwo\ in \cite{Lidman2012a},  $^h$ Mass from Finner et al. (submitted), 
        $^i$ \mfive from \cite{Brodwin2016}, $^\dagger$ Computed excluding CL J1226+3332.
         
	\end{tablenotes}
	\end{threeparttable}
	\label{table:piii_sample}
\end{table}

\section{Reduction}
\label{sec:piii_reduction}
 
We follow the same reduction procedure as in \citetalias{DeMaio2018} for the \red\ imaging of the high-redshift sample,  with one exception.  We do not apply a $\delta$-flat to each input image before drizzling the data into a final science image for each epoch of observations because this is a sub-dominant systematic for the current analysis. The pixel-to-pixel rms variation in the pipeline flat fields is at the level of \til0.5\%. This level of photometric uncertainty does not  
dominate our error because we measure both the sky and \bcgicl\ brightness over large areas and so the impact of local flatness variation is dramatically reduced. Any pixel-to-pixel correlations due to large-scale flatness variations across the detector are folded into our background measurement, which  quantifies  the uncertainty in the background as a function of position. More details on this method of constraining the flatness and background uncertainties are provided below. 

We drizzle each epoch of data separately and then median combine all epochs of a given cluster into a final science image. This  separate treatment of observations by epoch allows us to constrain the systematic variation in the background as a function of observation date. Of the seven high redshift clusters only the \idcs\ field includes foreground stars that are projected within 100~kpc of the BCG. These two stars are fairly faint ($\sim 19-20$ mag in F160W), and   masking is sufficient to ensure that light from the extended wings of the PSF does not significantly impact the observed \bcgicl\ luminosity.

Finally, we perform source masking and background determination as in \citetalias{DeMaio2018}. We identify sources  with Source Extractor \citep{SEx} and we mask all sources beyond 7\arcsec\ of the BCG to 3 times the semi-major and semi-minor axis of the Source Extractor catalogs. Within 7\arcsec\ we manually extend all mask radii to ensure that no remaining  light from other stars and galaxies contributes to the \bcgicl\ flux. 

\begin{figure*}
\begin{minipage}{0.32\textwidth}
    \includegraphics[width=\textwidth,trim={0 0.9cm 0 0},clip]
    {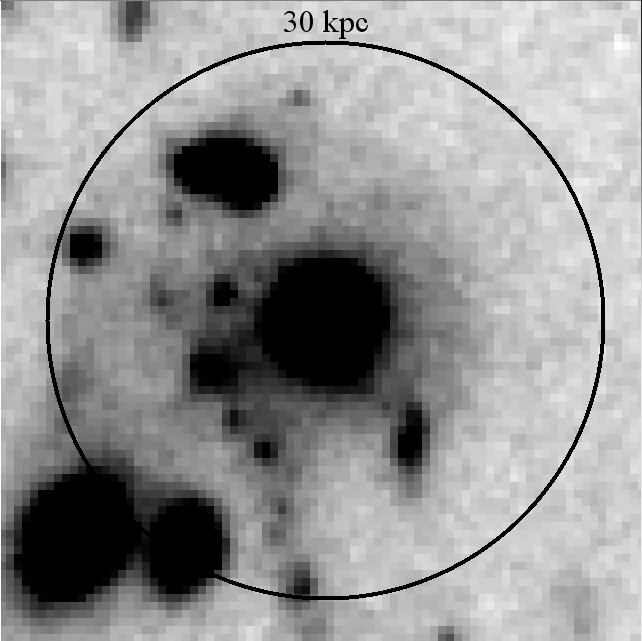}
\end{minipage}
\hfill
\begin{minipage}{0.32\textwidth}
    \includegraphics[width=\textwidth,trim={0 0.9cm 0 0},clip]
    {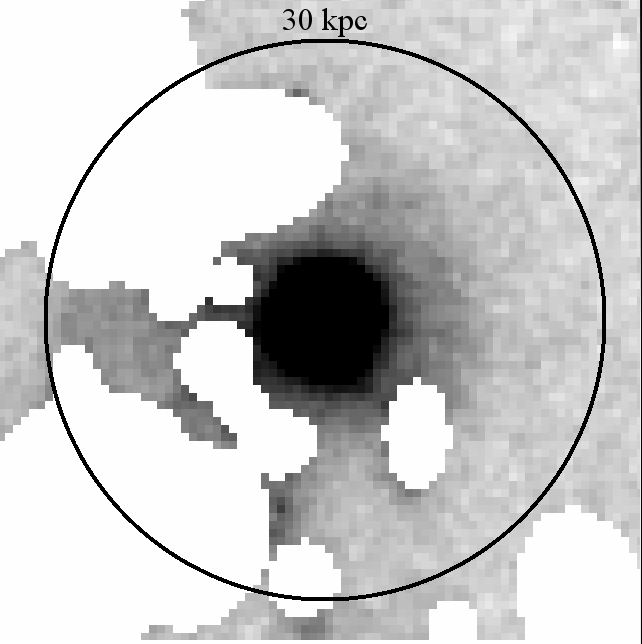}
\end{minipage}
\hfill
\begin{minipage}{0.32\textwidth}
\includegraphics[width=\textwidth,trim={0 0.9cm 0 0},clip]
{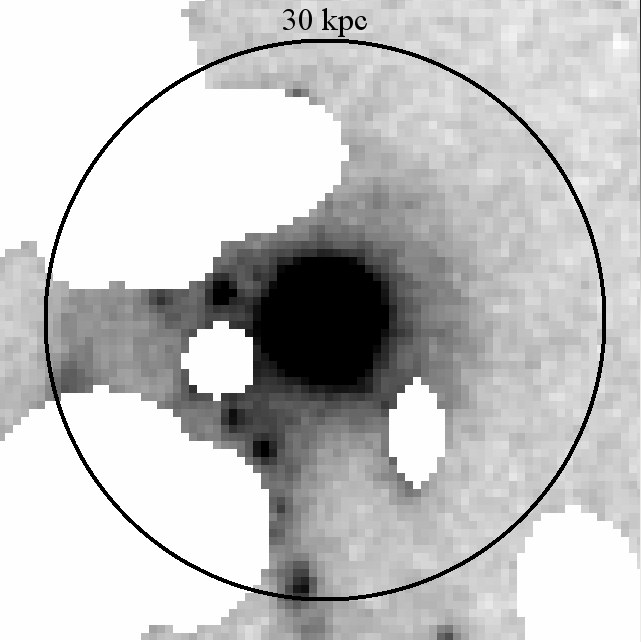}
\end{minipage}
\\* \vspace{1ex}
    A \hfill B \hfill C \\*
    \caption[Effects of masking differences on measured luminosities.]
    {We test the uncertainty on the measured luminosity due to differences in masking at the cluster core, using \sparcsonezero\ as an example. 
    A) Inner 30~kpc of \sparcsonezero\ in \red, unmasked.
    B) In the more severe masking regime we mask all pixels that are compact and elevated above any diffuse structure. Thus, for \sparcsonezero's 'beads on a string' morphology \citep{Webb2015} all bright knots (beads) are masked. 
    C) In the less restrictive masking regime we leave pixels unmasked if they are clearly associated with a distinct tidal feature or are irregular in shape and deeply embedded in the bright BCG envelope.
}
\label{fig:masking_diff}
\end{figure*}

The cluster cores of \xdcp\ and \sparcsonezero\ are clear ongoing mergers. For these two systems we test how differences in masking at the cluster core affect the measured luminosity within $r<100$~kpc by comparing flux measurements from a series of differently masked images. 
At one extreme we mask all pixels that are associated with compact, high surface-brightness features. For example, in the case of \sparcsonezero's 'beads on a string' morphology \citep{Webb2015} all bright knots (beads) are masked. In the less restrictive masking regime we leave pixels unmasked if they are clearly associated with a distinct tidal feature or are irregular in shape and deeply embedded in the bright BCG envelope. As an example, we show an unmasked image and both masking schemes for \sparcsonezero\ in Figure \ref{fig:masking_diff}. For \sparcsonezero\, differences in masking affect the total measured luminosity within 100~kpc by only 4\%. For \xdcp\, the difference in the measured \bcgicl\ luminosity within 100~kpc due to these differences in masking is 10\%. We use the less-extensive masking schemes throughout our
analysis and incorporate the uncertainty in flux due to  masking difference into the quoted luminosity and stellar mass measurement uncertainties.

After masking, we next determine the background level and uncertainty.
We first excise the inner 250~kpc of the cluster from the image. This excision ensures any bias in the sky determination due to the \bcgicl\ is negligible. The remaining unmasked pixels are divided into twenty-four $15^\circ$ wedges, centered on the BCG. We find the median of each wedge after an iterative 3$\sigma$ clipping  of pixel intensities inside each wedge.  The final background for the image is the mean of all wedge values. We define the scatter in wedge values to be the background uncertainty. Assessing the position-dependent variation in the measured background also probes the variation in flatness across the detector. Therefore, the final uncertainty on the background used throughout these analyses folds both flatness and background variation into our \bcgicl\ luminosity measurements.

Cosmological dimming makes measuring the \bcgicl\ component of clusters at high redshift a challenge. Solely due to this effect, the highest redshift cluster in our sample ($z=1.75$) has an observed surface brightness that is more than 10 times fainter than that of our average intermediate redshift cluster. However, the high redshift clusters have a significantly smaller angular size ($1.5\times$ smaller at $z>1$ than at $z=0.4$). The larger area available for background measurement  
allows us to make more precise background measurements and, in part, compensates for the observational challenge posed by cosmological dimming and allows us to derive \red\ surface profiles out to \til100~kpc for the $z>1$ clusters. 

\begin{figure}
    \centering
    \includegraphics[width=\columnwidth]{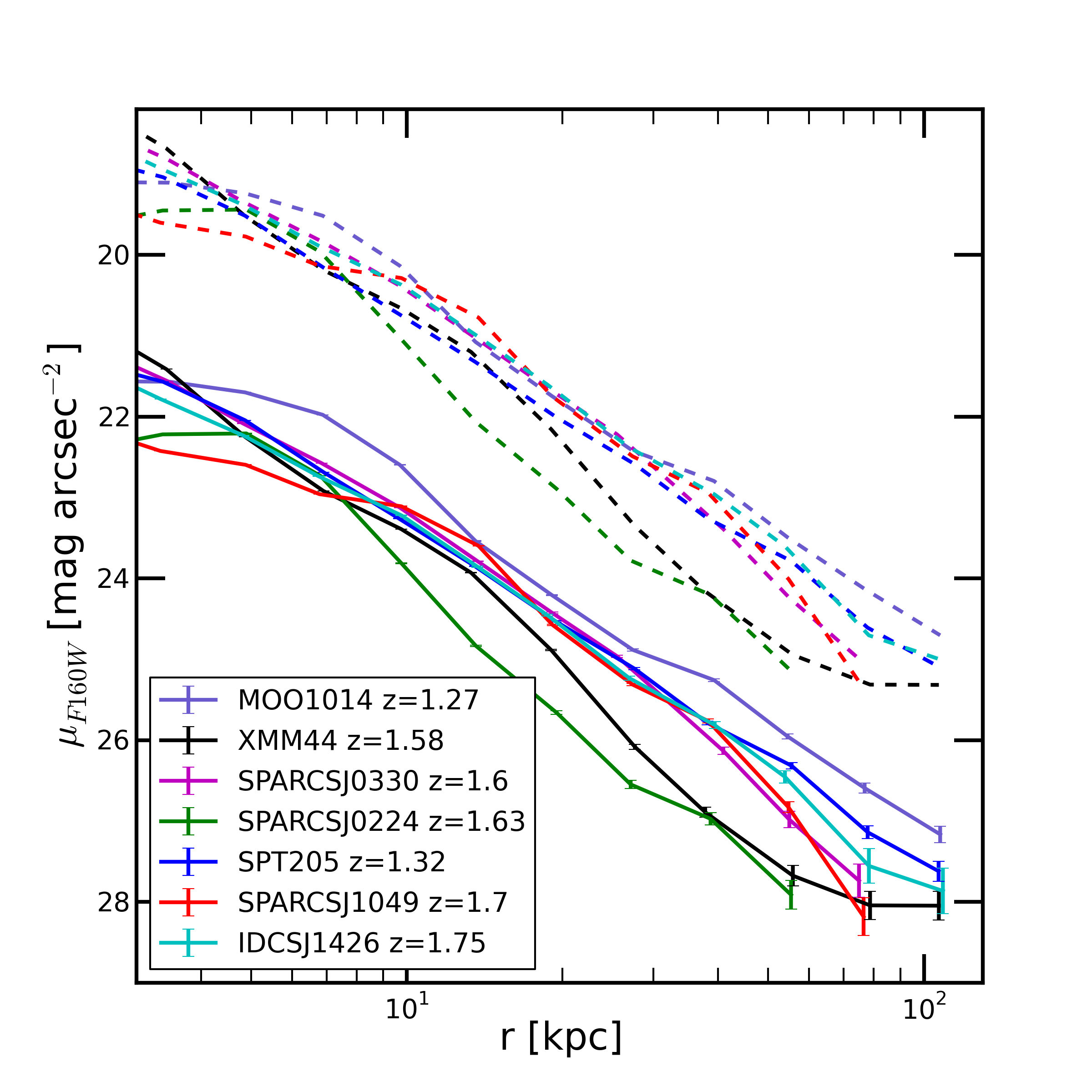}
    \caption[\red\ surface brightness profiles out to 100~kpc for the high redshift clusters.]
    {\red\ surface brightness profiles out to 100~kpc for the high redshift ($z>1$) clusters. Solid lines are observed profiles and dashed lines represent surface brightness profiles that have e+k and cosmological dimming corrections applied. The e+k corrections are derived from a \citetalias{BC03} SSP model with Chabrier IMF, $z_f=3$, and solar metallicity. Surface brightness profiles are terminated when the uncertainty reaches $>0.3$ \sbu. }
    \label{fig:sbprofs}
\end{figure}

Surface brightness profiles are produced by binning the \red\ images in radial bins with logarithmic bin widths of $d\log$($r$[kpc])$=0.05$ and 0.15. In Figure \ref{fig:sbprofs} we show the \red\
$d\log$($r$[kpc])$=0.15$ surface brightness profiles for the 7 high redshift clusters. To produce the evolution and passband (e+k) corrections for each system, we use a \citetalias{BC03} simple stellar population (SSP) model with Chabrier initial mass function (IMF) \citep{Chabrier2003}, formation redshift of $z_f=3$, and solar metallicity,  as in \citetalias{DeMaio2018}. The observed profiles  for each system are shown in solid lines and profiles that have been e+k corrected and cosmological dimming-corrected to $z=0$ are shown in dashed lines.

\section{Results}
\label{sec:results}
\subsection{The \Mstell\ $-$ \mfive\ Relation}
\label{sec:sm_content}

We convert from the observed luminosities of the \bcgicl\ to stellar masses by applying a  mass-to-light ratio  for the same stellar population model as in \S \ref{sec:piii_reduction}. All  photometric data (GZZ05, Paper II, and high-redshift) are converted using the same stellar population synthesis models.  In this conversion we do not attempt to account for gradients in M/L. We refer the reader to \citetalias{DeMaio2015} for a detailed discussion of metallicity gradients in the \bcgicl. Age gradients can also induce M/L gradients, but are expected to be small outside the central core of the BCG.

The luminosity and stellar mass in apertures of 10, 50, and 100~kpc are  given  for the high-redshift sample in Table \ref{table:sm_lum}. While it would also be of interest to measure these quantities within fixed fractions of \rfive, we cannot do so because of the restricted radial extent of our data coupled with the large range of \rfive,  which make it difficult to define $r/$\rfive\ apertures applicable to the full sample.\footnote{ \rfive\ is also a less robust quantity for comparison with simulations since it is derived from the X-ray data rather than directly observed.  Simulations must therefore adequately reproduce the X-ray data and associated uncertainties to avoid systematic bias between the simulations and observations.}
Values within fixed physical apertures thus provide more robust observable quantities.

\begin{figure}
\centering
    \includegraphics[width=\columnwidth]{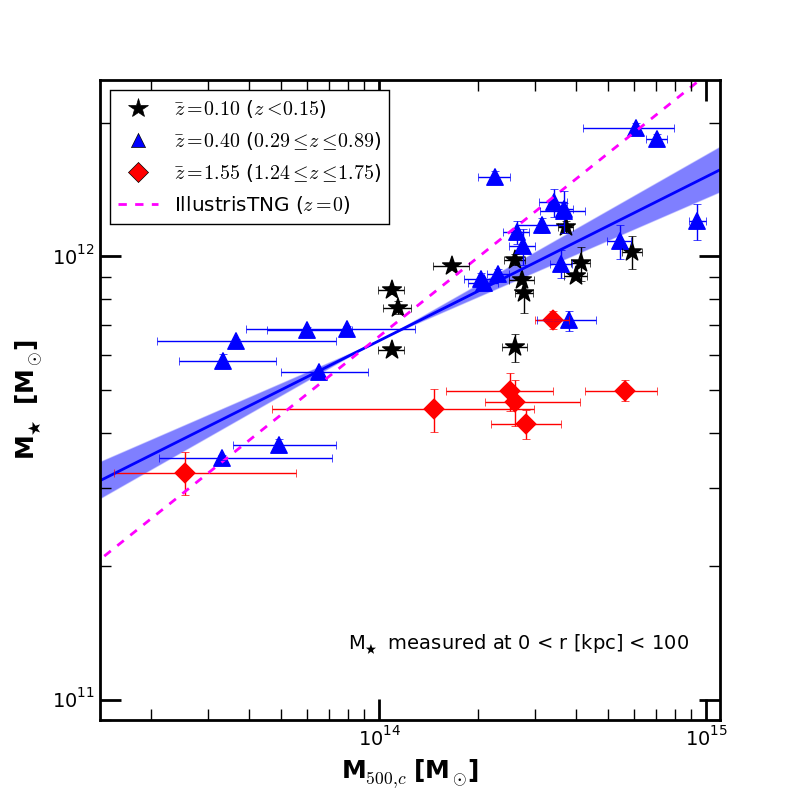}
    \caption[Stellar mass of the \bcgicl\ within $r<100$~kpc.]
    {\bcgicl\ stellar mass at $r<100$~kpc as a function of \mfive\ for the complete sample. Black stars are systems from \citetalias{Gonzalez2013a} with $z<0.15$, blue triangles are from \citetalias{DeMaio2018} (intermediate-redshift sample),  
    and red diamonds represent the high-redshift cluster sample. 
    The solid blue line represents the best-fit relation to the \bcgicl\ content within 100~kpc of the intermediate redshift sample with
    \logMstell$=(0.37\pm0.05)$\logmfivenorm $+ (11.92\pm0.02)$, with the shaded region corresponding to the 1$\sigma$ confidence region for the relation. The low-redshift sample of \citetalias{Gonzalez2013a} falls near this relation while the high-redshift sample on average exhibits lower \bcgicl\  stellar masses  at \mfive$>10^{14} M_\odot$. For comparison, we also show the $z=0$ relation derived for the IllustrisTNG simulations by \citet{pillepich2018}, which is slightly steeper than the observed relation but has a similar normalization. All stellar masses are derived at the cluster redshifts assuming passive evolution, with $z_f=3$ and solar metallicity, as described in the text.
    }
\label{fig:sm_m500_100}
\end{figure}

In Figure \ref{fig:sm_m500_100} we show the stellar mass of the \bcgicl\ within 100~kpc as a function of \mfive. In \citetalias{DeMaio2018} we found that the stellar mass within 100~kpc goes as \Mstell$\propto$\mfive$^{\alpha}$, with $\alpha=0.37\pm0.05$ for our sample of 23 clusters with $\bar z =0.4$ and a range in mass from 3\tenthirteen\ $-$ 9\tenfourteen\ \Msun. From Figure \ref{fig:sm_m500_100}, we see that the sample of low-redshift clusters (black stars) also lies on this relation, shown in blue. For the high-redshift systems, we find smaller total stellar masses within 100~kpc at a given \mfive for clusters more massive than $10^{14} M_\odot$.

The IllustrisTNG simulations provide an interesting point of comparison. We overplot in Figure \ref{fig:sm_m500_100} the \Mstell$-$\mfive\ relation derived by \citet{pillepich2018} at $r<100$~kpc for IllustrisTNG. Those authors commented that at $r<30$~kpc the slope derived from the simulations was too steep compared to the data from \citet{Kravtsov2014}, but had a similar normalization at \mfive$=10^{14} M_\odot$. We find a similar result compared to \citet{pillepich2018}, but now for data extending to $r \sim 100$~kpc. The slope from the simulations is somewhat steeper ($\alpha=0.59$) than the slope from our \citetalias{DeMaio2018},  but the normalization at \mfive$=10^{14}$~kpc is similar. As noted in \citet{pillepich2018}, a similar mismatch at the highest masses is seen in other simulations \citep[e.g.][]{ragonefigueroa2013,hahn2017,bahe2017}.

\begin{figure}
\includegraphics[width=\columnwidth]{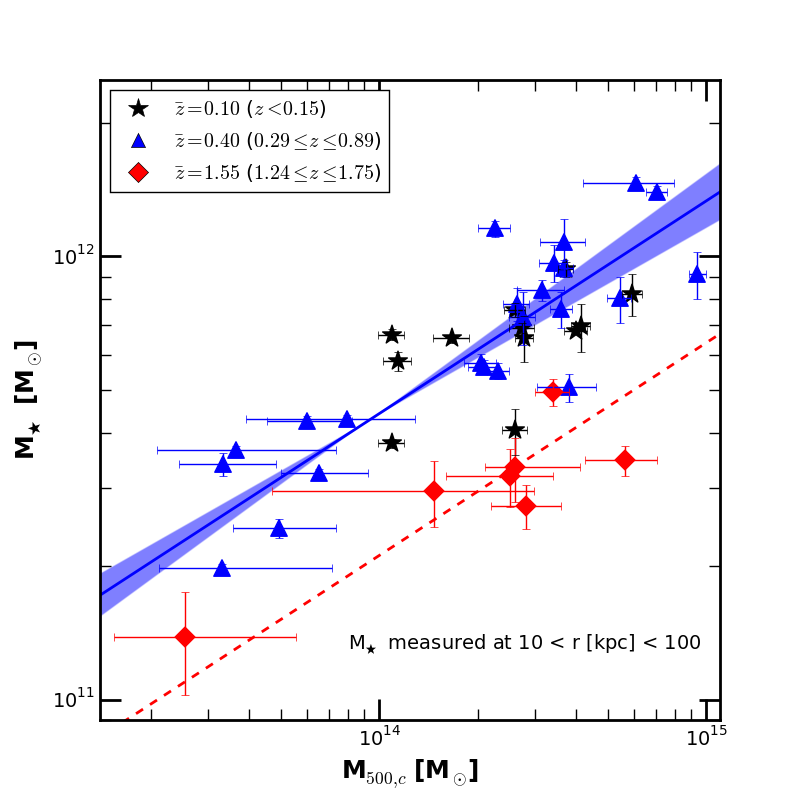}
    \caption[ Stellar mass of the ICL within $10<r<100$~kpc.]{Stellar mass within $10<r<100$~kpc as a function of \mfive.   Markers and colors are as in Figure \ref{fig:sm_m500_100}. The solid blue line is the best-fit to the intermediate redshift sample,   also as in Figure \ref{fig:sm_m500_100}. Excluding the stellar mass in the central 10~kpc produces a steeper relationship with \mfive,
    \logmstell$=(0.48\pm0.06)$\logmfive$+(11.79\pm0.02$), indicating that 
the central BCG represents a larger fraction of the total \bcgicl\ light inside 100~kpc in groups 
relative to clusters.  The high redshift clusters are consistent with following the same functional relation as at lower redshifts, but shifted to lower \Mstell\ by a factor of  $2.08\pm0.21$. The best-fit relation for the high-redshift sample is shown in dashed red, assuming the same slope as for the  intermediate redshift sample. 
}
    \label{fig:sm_m500_ring}
\end{figure}

In Figure \ref{fig:sm_m500_ring} we show the same clusters, but now excluding the inner 10~kpc from the mass measurements, as the stellar mass is dominated by the central BCG at these radii. We perform an orthogonal distance regression fit to the intermediate redshift data with the relation
\begin{equation}
\log \textrm{M}_{\star}=\alpha \log\left( \frac{M_{500,c}}{2\times10^{14} M_\odot} \right) + \beta .
\label{eqn:mstell}
\end{equation}
In this equation  the normalization of the relation, $\beta$, is the log of the stellar mass at $10<r<100$~kpc for a cluster with \mfive$=$2\tenfourteen\ \Msun. For $10<r<100$~kpc, the $\log~$\Mstell\ $-~\log~$\mfive\ relation for the intermediate sample has a slightly steeper slope, $\alpha=0.48\pm0.06$, than when the central 10~kpc are included. This fit is derived excluding the cluster CL J1226+3332 at $z=0.89$. CL J1226+3332 is the only cluster in any of the subsamples that lies between $z=0.6$ and $z=1.2$. By excluding it we narrow the redshift window over which the intermediate-redshift relation is determined.

The data for the high-redshift sample are consistent with having the same slope, but the large uncertainties and limited number of low-mass clusters preclude a meaningful, independent determination of the slope for this sample.
We therefore in the next section assume a redshift-independent slope of $\alpha=0.48$ and focus upon constraining evolution in the normalization $\beta$.

\begin{figure}
\centering
    \includegraphics[width=\columnwidth]
    {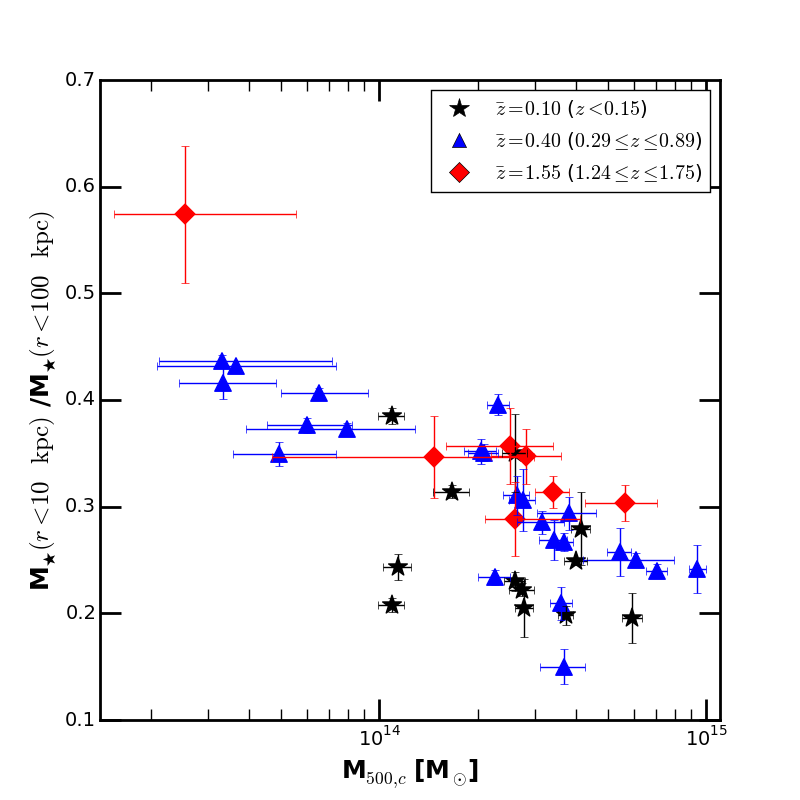}
    \caption[Fraction of stellar mass of the \bcgicl\ within 10~kpc relative to the total within 100~kpc.]
    {The fraction of the stellar mass within the central 10~kpc relative to the total within 100~kpc, plotted as  a function of \mfive\ for the complete sample. Systems at all redshifts show a trend of lower central concentration with increasing \mfive, with no significant shift in this relation betwen samples.}
\label{fig:sm_m500_totmassratio}
\end{figure}

\begin{table*}
\caption{Luminosity and Stellar Mass for High-redshift Sample}
\begin{threeparttable}
\begin{tabular}{cccccccc}
    \hline
    Cluster & z & L (r$\le$10 kpc) & L (r$\le$50 kpc) & L (r$\le$100 kpc) & M$_\ast$ (r$\le$10 kpc) & M$_\ast$ (r$\le$50 kpc) & M$_\ast$ (r$\le$100 kpc) \\
 &  & [$10^{11} L_\odot$] & [$10^{11} L_\odot$] & [$10^{11} L_\odot$] & [$10^{11} M_\odot$] & [$10^{11} M_\odot$] & [$10^{11} M_\odot$] \\
 \hline
MOO1014 & 1.24 & 3.94$\pm$0.01 & 9.24$\pm$0.27 & 12.43$\pm$0.60 & 2.21$\pm$0.01 & 5.20$\pm$0.09 & 6.99$\pm$0.34 \\
SPT205 & 1.32 & 2.67$\pm$0.01 & 6.60$\pm$0.23 & 8.79$\pm$0.48 & 1.51$\pm$0.01 & 3.74$\pm$0.07 & 4.98$\pm$0.27 \\
XMM44 & 1.58 & 2.49$\pm$0.69 & 5.63$\pm$0.93 & 7.18$\pm$0.94 & 1.46$\pm$0.24 & 3.29$\pm$0.32 & 4.20$\pm$0.55 \\
SPARCSJ0330 & 1.60 & 2.67$\pm$0.01 & 6.62$\pm$0.37 & 7.71$\pm$0.85 & 1.57$\pm$0.01 & 3.88$\pm$0.13 & 4.52$\pm$0.50 \\
SPARCSJ0224 & 1.63 & 3.17$\pm$0.01 & 4.98$\pm$0.28 & 5.53$\pm$0.62 & 1.87$\pm$0.01 & 2.93$\pm$0.10 & 3.25$\pm$0.36 \\
SPARCSJ1049 & 1.70 & 2.98$\pm$0.34 & 7.76$\pm$0.60 & 8.33$\pm$0.81 & 1.77$\pm$0.12 & 4.63$\pm$0.21 & 4.97$\pm$0.48 \\
IDCSJ1426 & 1.75 & 2.25$\pm$0.01 & 6.06$\pm$0.38 & 7.79$\pm$0.92 & 1.36$\pm$0.01 & 3.66$\pm$0.14 & 4.70$\pm$0.56 \\
\hline
\end{tabular}
\begin{tablenotes}
    \item  Stellar masses are derived using the stellar population model described in section \ref{sec:sm_content}. 
    Quoted uncertainties on the stellar mass do not reflect systematic errors associated with the mass to light conversion,  including the choice of IMF and stellar population age and metallicity. 
     These uncertainties also do not account for any gradients in the M/L ratio of the  \bcgicl,  as seen in \citetalias{DeMaio2015}.
\end{tablenotes}
\end{threeparttable}
\label{table:sm_lum}
\end{table*}

\subsection{Evolution in the \Mstell\ $-$ \mfive\ Relation and \bcgicl\ Stellar Mass Growth}

On its surface, the coincidence of the low and intermediate redshift points in Figures \ref{fig:sm_m500_100} and \ref{fig:sm_m500_ring} suggests little to no evolution in the \bcgicl\ since  $\bar{z} = 0.4$.  
However, semi-analytic models and numerical simulations typically predict substantial late-time growth of the ICL. For example, the semi-analytic models of \citet{Contini2013a,contini2018}, which are based upon the Millennium simulations, predict that the ICL has doubled in stellar mass since $z=0.5$. Several recent observational studies concur, suggesting that this growth is driven by the addition of stars from mergers and stripping events \citep{Burke2012a, Zhang2016}. In contrast, the lack of evolution we observe between the local and intermediate-redshift \Mstell\ $-$ \mfive\ relations in Figures \ref{fig:sm_m500_100} and \ref{fig:sm_m500_ring}  suggests that late \Mstell\ growth within $100$~kpc is modest. 

One way out of this conclusion is to posit that \Mstell\ growth is matched by growth in \mfive\ so that clusters evolve along the \Mstell\ $-$ \mfive\ relation.  However, the expected modest mass growth in \mfive, $\sim$20\% from $z = 0.4$ to $z=0.1$, limits the \Mstell\ growth in this scenario to $<$ 10\%. Therefore, reconciliation with the theoretical expectation is only possible if any substantial late time ICL growth occurs at $r>100$~kpc, a radial regime that is not constrained by our data. \footnote{We note that in this regard simulations have an advantage relative to semi-analytic models, as the latter lack spatial information on galaxy stellar distributions.
}

Although there is little if any evolution in the normalization $\beta$ of the relation from $\bar{z} = 0.4$ to $\bar{z} =0.1$, there is evolution prior to  $\bar{z}= 0.4$. To quantify that evolution, we fit each of our three redshift sub-samples with equation \ref{eqn:mstell} using a fixed slope of $\alpha=0.48$, as found for the intermediate-redshift sample at $10<r<100$~kpc. For the intermediate-redshift sample, we again exclude the cluster CL J1226+3332 from the fit.

We present our $\beta$ values for each redshift bin in Table \ref{tab:bevo_10-100}. As expected from Figure \ref{fig:sm_m500_ring}, we see that the  low-redshift and intermediate-redshift samples have statistically consistent normalization ($\beta = 11.77\pm0.03$  and $\beta = 11.79\pm0.02$, respectively). For the high-redshift sample, the normalization is significantly different ($\beta=11.47\pm0.04$). Thus, between the high-redshift ($\bar{z}=1.55$) and intermediate-redshift  ($\bar{z}=0.4$) samples, the normalization of the \Mstell$-$\mfive\ relation  evolves by a factor of $2.08\pm0.21$ ($\Delta\beta=0.32\pm0.06$) for \Mstell\ measured within $10<r<100$~kpc.
At lower redshift, any growth in the \bcgicl\ stellar mass within 
$10<r<100$~kpc must correspond to evolution along the observed \Mstell$-$\mfive\ relation. 

Our results indicate that the epoch between the high- and intermediate-redshift samples ($z\simeq0.6-1.2$) is a key period for \bcgicl\ development within 100~kpc. The observed lack of evolution in the \Mstell$-$\mfive\ relation at low-redshift is correctly predicted by   \citet{pillepich2018}. These authors find no evolution in IllustrisTNG since $z=1$, which coupled with our data suggests that the bulk of the evolution may occur rapidly just before this epoch. It is  important for future studies to add clusters at $z\sim 1$ to bridge the gap between our intermediate- and high-redshift samples and thus better map the redshift evolution of this relation.

\begin{table}
    \caption{\Mstell-\mfive\ Normalization ($\beta$) of the \Mstell-\mfive\ Relation }
	\begin{center}
	\begin{threeparttable}

		\begin{tabular}{p{0.2\columnwidth}cp{0.3\columnwidth}cp{0.3\columnwidth}}
		    \hline
            $\bar{z}$ & $z$ range & $\;\;\;\;\;\;\;\;\beta$ \\
    		\hline
    		
            0.10 & [0.0578,0.1237] & $11.77\pm0.03$\\
            0.40 & [0.288,0.57] &$11.79\pm0.02$\\
            1.55 & [1.24,1.75] & $11.47\pm0.04$\\
            \hline
		\end{tabular}
	\begin{tablenotes}
        \item  Normalization
         ($\beta$) of the best-fit \Mstell-\mfive\ relation for all redshift bins with an assumed slope equal to that of best-fit to the  intermediate redshift  sample. The high-redshift sample shows a lower $\beta$  
        compared to the samples the lower redshift samples, which have the same $\beta$  
        within uncertainties. The data thus imply that there is a transition from rapid \bcgicl\ growth at $10<r<100$~kpc at early times to minimal growth at these radii subsequently.
	\end{tablenotes}
	\end{threeparttable}
	\end{center}
    \label{tab:bevo_10-100}
\end{table}

\subsection{Growth of the \bcgicl\ in a Typical Cluster}
 
We are not directly observing an evolutionary sequence between epochs, but rather measuring \Mstell\  over a similar range in \mfive\ at three epochs. To constrain the growth of the \bcgicl, we must adopt the results from models of structure formation that describe how clusters grow in mass between the three epochs.

For a \citet{tinker2008} mass function, evolution of the cluster mass function results in an increase in \mfive\ of roughly a factor of 4-5 from $z=1.5$, which is approximately the mean redshift of our high-redshift sample, to $z=0$. For example, a galaxy group with \mfive=5\tenthirteen\ \Msun\ at $z=1.5$ will grow to \mfive$\simeq$ 2\tenfourteen\ \Msun\ by $z=0$  \citep[e.g. \url{hmfcalc.org},][]{hmfcalc2013}.

One might na\"ively expect similar growth in \Mstell. However, as emphasized by \citet{diemer2013a,diemer2013b}, this increase in \mfive\ is not all due to true physical growth of the cluster. Rather, much of the increase is due to pseudo-evolution arising from the definition of \mfive\ relative to the critical density ($\rho_c$). As $\rho_c$ decreases with time, the radius within which \mfive\ is measured increases, leading to a corresponding increase in \mfive. After 
correcting for this pseudo-evolution via the approach in \citet{diemer2013a}, the growth for a cluster with \mfive=5\tenthirteen\ \Msun\ at $z=1.5$ is only a factor of 1.4 from $z=1.5$ to $z=0$.

Keeping this factor in mind, we now consider the inferred stellar mass growth for this cluster, using the observed \Mstell$-$ \mfive\ relation. If the  \Mstell\ $-$ \mfive\ relation does not evolve, then the factor of four growth in \mfive\ corresponds to a factor of 1.9 ($4^{\alpha}$ for $\alpha=0.48$) growth in \Mstell, the stellar mass content at $10<r<100$~kpc. Including evolution of the normalization $\beta$
with redshift, we find that at $10<r<100$~kpc the stellar mass of the \bcgicl\ grows by a factor of 3.8 -- 1.9$\times$ from mass growth along the sequence and $2\times$ associated with the increase in $\beta$. The bulk of this growth occurs during the epoch between the high- and intermediate-redshift samples. It is during this epoch when we see $\beta$ evolve, and when \mfive\ is increasing most rapidly.

We now place these results within the context of previous measurements.
The level of evolution in the ICL content here---a factor of $\sim3.8$ since $\bar{z}=1.55$---is similar to the factor of $2-4$ increase in the \bcgicl\ light at $\mu_J>22$ \sbu, within \rfive, and since $z=1$ measured by \cite{Burke2012a}. This amount of growth is also consistent with observed evolution in stellar mass density profiles \citep[see Figure 7 of][]{vanderburg2015}. It is, however, lower than the  growth of 90\% since $z=1$ predicted by \citet{Contini2013a,contini2018}.
As we discussed previously, this contradiction can be resolved by invoking significant growth at radii beyond 100~kpc. We infer that  with our data we are witnessing  substantial  growth of the \bcgicl\ at $r < 100$~kpc between $\bar{z} = 1.55$ and 0.4. At lower redshifts, any growth must occur outside of our observational aperture. It is important that simulations address the radii at which growth is occurring as a function of redshift.

\subsection{Evolution of the \bcgicl\ Spatial Distribution}

While we cannot probe the evolution of the \bcgicl\ beyond 100~kpc, we can investigate whether we see evidence of inside-out growth over the radial range of our data. Specifically, we now consider whether there is relative evolution  between the inner 10~kpc and the annulus at $10<r<100$~kpc. Figure \ref{fig:sm_m500_totmassratio} shows the fraction of the total stellar mass within 100~kpc that lies within the central 10~kpc. There is no statistically significant evolution in this relation 
between the high-redshift sample  and the intermediate-   and low-redshift   samples. We also obtain similar results for larger central apertures (e.g. 20~kpc). 

We conclude that at fixed cluster mass, the concentration of the stellar mass is constant across redshift. However, since any individual cluster will continue to gain mass over time, this conclusion does not mean that concentration of stellar mass for a single cluster stays fixed. Rather, as a cluster grows, we expect its stellar mass concentration to evolve along the relation shown in Figure 6. For a given cluster the \bcgicl\ will become less concentrated over time, and a higher proportion of stellar mass that lies outside the central 10~kpc. This  accretion of stars with ever-larger orbits in the dark matter potential is inside-out growth.

The paradigm of inside-out growth has previously been invoked for massive ellipticals.  Compact, dense stellar cores form very early ($z>2$) and subsequent stellar mass growth occurs at larger radii \citep{vanDokkum2010, Patel2013, vanderBurg2014, Bai2014, Hill2017, VandeSande2013}. The \bcgicl\ constitutes the extreme mass end of the elliptical population, and it is therefore natural to expect similar properties and stellar mass growth histories. \cite{Patel2013} found that 50\% of the light within 100~kpc of massive galaxies at $z\sim1.5$ is contained outside of 10~kpc (see also \citealt{Hill2017} for similar conclusions), similar to the fraction that we observe for the \bcgicl\ in the lowest mass groups here (Figure \ref{fig:sm_m500_totmassratio}). Our findings quantify this growth for the \bcgicl\ and are not qualitatively different than what one would have expected given our understanding of the growth of massive ellipticals.

\section{Conclusions}
\label{sec:piii_conclusion}

By combining three samples of galaxy groups and clusters spanning a wide range in redshift ($0.03<z<1.75$) and mass (\mfive$=2.5$\tenthirteen$-$8\tenfourteen\ \Msun), we measure the growth of the stellar mass of the \bcgicl, \Mstell, within $10<r<100$~kpc over the last ten Gyr. In particular:
\begin{enumerate}
 \item  We derive a best-fit relation between \Mstell\ at $10<r<100$~kpc and \mfive\  of the form \logMstell$=\alpha$\logmfivenorm$ + \beta$  for our intermediate-redshift sample ($\bar{z}=0.4$) -- the sample for which we have the best data. We find a slope $\alpha=0.48\pm0.06$ and a normalization $\beta=11.79\pm0.02$. The slope of this relation  is steeper than that obtained for $r<100$~kpc in \citetalias{DeMaio2018} ($\alpha=0.37\pm0.05$). The difference comes from excluding here 
 the inner 10~kpc, where the BCG dominates the stellar mass.

 \item 
 Fixing the slope to $\alpha=0.48$ for our other redshift samples, we quantify how the normalization, $\beta$, of the \Mstell$-$\mfive\ relation evolves. For a cluster with \mfive$=$ 2\tenfourteen\ \Msun, 
 the \bcgicl\ stellar mass within $10< r< 100$~kpc is given by 
 $\beta=11.77\pm0.03$ for the low-redshift sample  ($\bar{z}=0.10$),  $11.79\pm0.02$ for the intermediate-redshift sample  ($\bar{z}=0.40$), and $11.47\pm0.04$ for the high-redshift sample  ($\bar{z}=1.55$). 
 There is a factor of $2.08\pm0.21$ increase in the stellar mass between the high- and intermediate-redshift samples at a fixed \mfive.

 \item  We consider the overall increase in \Mstell\ for a typical cluster since $z=1.5$ using the observed \Mstell$-$\mfive\ relations in conjunction with theoretical predictions for the evolution of the halo mass function.  For a cluster with \mfive$\simeq$ 2\tenfourteen\ \Msun\ at $z=0$, the \bcgicl\ stellar mass within $10 < r < 100$~kpc  with increases by a factor of $\sim3.8$ since $z=1.5$. This growth in the \bcgicl\ stellar mass exceeds the physical mass increase of the cluster, which is only a factor of 1.4 after correcting for the pseudo-evolution of \mfive\ arising from the changing critical density $\rho_c$.

    \item We find evidence for inside-out mass growth of the \bcgicl\ at $r<100$~kpc over the past ten Gyrs. We characterize the spatial concentration of the \bcgicl\ using the ratio of the \bcgicl\ stellar mass within 10~kpc to the total \bcgicl\ stellar mass within 100~kpc. We observe an anti-correlation between the \bcgicl\ concentration and \mfive\ that is independent of redshift. Consequently, as a  cluster grows in \mfive,  it must preferentially gain stellar mass at $r>10$~kpc to remain on the observed relation.

\end{enumerate}

We interpret the above results in terms of inside-out growth of the \bcgicl\ that eventually progresses to radii beyond those encompassed in our measurement aperture. We  directly observe that the  \bcgicl\  gains stellar mass at 10 to 100~kpc more rapidly than within the central 10~kpc. At early times, the stellar halo at $10<r<100$~kpc forms more rapidly than the physical mass growth of the cluster, as evidenced by the evolution of the normalization in the \Mstell$-$\mfive\ relation. At later times (since $\bar{z}=0.4$), there is little stellar mass growth at these radii. Our observational constraints thus require that continued \bcgicl\ growth must occur at even larger radii if theoretical models predicting significant late-time growth are correct. Tracing the evolution of the \bcgicl, especially beyond  $\sim$100~kpc, should be a priority for simulators, as future facilities such as {\it Euclid} and {\it WFIRST} will be able to measure the \bcgicl\ growth at those radii.

\section*{Acknowledgements}

We thank the referee, Emanuele Contini, for his careful review of this paper. We also thank Adam Muzzin for his helpful comments, and Kyle Finner and James Jee for providing a lensing mass for SpARCS J1049.
 We acknowledge support from the National Science Foundation
through grant NSF-1108957. Support for Program numbers 12634 and 12575 was provided by NASA through a grant from the Space Telescope Science Institute, which is operated by the Association of Universities for Research in Astronomy, Incorporated, under NASA contract NAS5-26555. AIZ acknowledges support from NSF grant AST-0908280 and NASA grant ADP-NNX10AD47G. The archival and Guest Observer data are based on
observations made with the NASA/ESA Hubble Space Telescope, which is operated by the Space Telescope Science Institute.

\bibliographystyle{mnras}
\bibliography{PaperInIInIII} 



\bsp	
\label{lastpage}
\end{document}